\documentclass[]{pasj01}

\begin{document} 
\Received{}
\Accepted{}

\title{AKARI and IRAS: From Beam Corrections to SEDs}

\author{David L. \textsc {Clements}\altaffilmark{1}}%
\altaffiltext{1}{Imperial College London, Blackett Lab., Prince Consort Road, London, SW7 2AZ, UK}
\email{d.clements@imperial.ac.uk}

\author{Michael \textsc{Rowan-Robinson},\altaffilmark{1}}
\email{m.rrobinson@imperial.ac.uk}

\author{Chris \textsc{Pearson}\altaffilmark{2,3,4}}
\altaffiltext{2}{RAL Space, CCLRC, Rutherford Appleton Laboratory, Chilton, Didcot, Oxfordshire OX11 0QX, UK}
\altaffiltext{3}{School of Physical Sciences, The Open University, Milton Keynes, MK7 6AA, UK}
\altaffiltext{4}{Oxford Astrophysics, Denys Wilkinson Building, University of Oxford, Keble Rd, Oxford OX1 3RH, UK}
\email{chris.pearson@stfc.ac.uk}

\author{Jose \textsc {Afonso}\altaffilmark{5,6}}%
\altaffiltext{5}{Instituto de Astrof\'{i}sica e Ci\^{e}ncias do Espa\c co, Universidade de Lisboa, OAL, Tapada da Ajuda, PT1349-018 Lisboa, Portugal}
\altaffiltext{6}{Departamento de F\'{i}sica, Faculdade de Ci\^{e}ncias, Universidade de Lisboa, Edif\'{i}cio C8, Campo Grande, PT1749-016 Lisbon, Portugal}
\email{jafonso@oal.ul.pt}

\author{Vianney \textsc {Labouteiller}\altaffilmark{7,8}}%
\altaffiltext{7}{Universit{\'e} Paris Diderot, AIM, Sorbonne Paris Cit{\'e}, CEA, CNRS, F-91191 Gif-sur-Yvette, France}
\altaffiltext{8}{IRFU, CEA, Universit{\'e} Paris-Saclay, F-91191 Gif-sur-Yvette, France}
\email{myravian@gmail.com}

\author{Duncan \textsc{Farrah},\altaffilmark{9, 10}}
\altaffiltext{9}{Department of Physics and Astronomy, University of Hawaii, 2505 Correa Road, Honolulu, HI 96822, USA}
\altaffiltext{10}{Institute for Astronomy, 2680 Woodlawn Drive, University of Hawaii, Honolulu, HI 96822, USA}
\email{dfarrah@hawaii.edu}

\author{Andreas \textsc{Efstathiou},\altaffilmark{11}}
\altaffiltext{10}{School of Sciences, European University Cyprus, Diogenes Street, Engomi, 1516, Nicosia, Cyprus}
\email{A.Efstathiou@euc.ac.cy}

\author{Josh \textsc{Greenslade},\altaffilmark{12}}
\email{j.greenslade14@imperial.ac.uk}

\author{Lingyu \textsc{Wang},\altaffilmark{13,14}}
\altaffiltext{11}{SRON Netherlands Institute for Space Research, Landleven 12, 9747 AD, Groningen, The Netherlands}
\altaffiltext{12}{Kapteyn Astronomical Institute, University of Groningen, Postbus 800, 9700 AV, Groningen, The Netherlands}
\email{L.Wang@sron.nl}



\maketitle

\begin{abstract}
There is significant scientific value to be gained from combining {\em AKARI} fluxes with data at other far-IR wavelengths from the {\em IRAS} and {\em Herschel} missions. To be able to do this we must ensure that there are no systematic differences between the datasets that need to be corrected before the fluxes are compatible with each other. One such systematic effect identified in the BSCv1 data is the issue of beam corrections. We determine these for the BSCv2 data by correlating ratios of appropriate {\em IRAS} and {\em AKARI} bands with the difference in 2MASS $J$ band extended and point source magnitudes for sources cross matched between the {\em IRAS} FSC, {\em AKARI} BSCv2 and 2MASS catalogs. We find significant correlations ($p<< 10^{-13}$ correlations indicating that beam corrections are necessary in the 65 and 90$\mu$m bands. We then use these corrected fluxes to supplement existing data in spectral energy distribution (SED) fits for ultraluminous infrared galaxies in the HERUS survey. The addition of {\em AKARI} fluxes makes little difference to the results of simple $(T,\beta)$ fits to the SEDs of these sources, though there is a general decrease in reduced $\chi^2$ values. The utility of the extra {\em AKARI} data, however, is in allowing physically more realistic SED models with more parameters to be fit to the data. We also extend our analysis of beam correction issues in the {\em AKARI} data by examining the Herschel Reference Sample, which have {\em Herschel} photometry from 100 to 500$\mu$m and which are more spatially extended than the HERUS ULIRGs. Thirty-four of the HRS sources have good {\em Herschel} SEDs and matching data from {\em AKARI}. This investigation finds that our simple 2MASS-based beam correction scheme is inadequate for these larger and more complex sources. There are also indications that additional beam corrections at 140 and 160$\mu$m are needed for these sources, extended on scales $>$1 arcminute.
\end{abstract}

\section{Introduction}

The thermal emission of cool dust, at temperatures of $\sim$20-60K, is a major constituent of the spectral energy distribution (SED) of all star forming galaxies. Normal spirals, like the Milky Way Galaxy, emit $\sim$ 30\% of their energy through cool dust emission (eg. \cite{2012MNRAS.427.2797D}) in the far-IR, at wavelengths peaking near to 100$\mu$m. Since the dust is heated through absorbing optical/UV emission from a galaxy's stellar population and/or AGN, this implies that roughly 30\% of the light from stars in a typical galaxy is absorbed by dust. Galaxies with higher star formation rates have a greater fraction of their energy output in the far-IR, with the most extreme objects in the local universe, the Ultraluminous IRAS Galaxies (ULIRGs with L$_{fir}>10^{12}$M$_{\odot}$) emitting over 90\% of their energy in the far-IR (e.g. \cite{1984Natur.309..430W}, \cite{1984ApJ...283L...1S}), with the vast majority of the energy generated by their starbursts obscured by dust. While the most luminous, ULIRG-class, objects are rare in the local universe, this population evolves rapidly with redshift, with their higher redshift equivalents contributing 3-5 orders of magnitude more to the co-moving luminosity density at $z>1$ than they do locally (\cite{2005ApJ...632..169L}), possibly becoming the dominant source of far-IR luminosity at the epoch of peak star formation around $z\sim2$ (\cite{p12}, \cite{2013A&A...553A.132M}, \cite{2013MNRAS.432...23G}). Understanding the nature of the far-IR emitting dust and the powers sources driving the far-IR emission in all galaxies, and especially in the most luminous ULIRG sources, is thus an important task for far-IR astronomy.

Since the dust SED peaks at wavelengths around $\sim$100$\mu$m, observations in the far-IR are required to determine the properties of the dust emission. Observations at wavelengths close to this peak are especially important for accurate temperature determination and to see if there are any signs that the dust obscuration might be optically thick (see eg. \cite{2011ApJ...743...94R}) and to search for evidence of multiple populations of dust at different temperatures (see eg.  \cite{2001MNRAS.327..697D}). Because far-IR observations are largely impossible from the surface of the Earth, this work requires data from space based observatories. {\em IRAS} observations have been the mainstay for this work for many years since it provides an all-sky survey at 12, 25, 60 and 100$\mu$m. Observations with {\em ISO} expanded this dataset and extended fluxes out to 170$\mu$m (\cite{2003AJ....125.2361B}, \cite{2007A&A...466..831S}) whilst those with {\em Spitzer} went much deeper and added 70 and 160$\mu$m fluxes (eg. \cite{2003PASP..115..928K}). However, {\em ISO} and {\em Spitzer} covered only a small fraction of the sky, so relatively limited samples of targets are available that have these data. More recently the large area {\em Herschel} surveys have covered between them $\sim$1000 sq. deg. of sky at 250, 350 and 500$\mu$m to 1$\sigma$ sensitivities of a few mJy (eg. \cite{2012MNRAS.424.1614O}, \cite{2010PASP..122..499E}). Samples of specific sources of interest, such as ULIRGs (eg. \cite{2013ApJ...776...38F}) were additionally targeted at shorter wavelengths adding flux points at 70, 100 and 160$\mu$m.

However, two things are lacking from the compendium of data currently available to far-IR astronomers: improved coverage of the wavelength region between $\sim$90 and 160$\mu$m where the SED peaks, which is important in determining optical thickness and/or the presence of dust at multiple temperatures; an all sky survey comparable to {\em IRAS} but adding data at wavelengths longer then 100$\mu$m. The {\em AKARI} mission provides both of these requirements, so it potentially has a key role to play in the analysis of far-IR SEDs of galaxies in the local universe, and especially the local ULIRGs. However, in order for this potential to be realised, we must make sure that the {\em AKARI} fluxes can be accurately combined with data from {\em IRAS}, {\em Herschel} and other space missions with any systematic photometric offsets corrected.

In this paper we search the {\em AKARI} Bright Source Catalog version 2 (BSCv2 \cite{y18}) for evidence that beam corrections are needed to account for any far-IR flux missed from extended sources, and then combine the corrected fluxes with other far-IR data to examine the far-IR SEDs of local ULIRGs from the HERUS survey (\cite{2013ApJ...776...38F}, \cite{2016ApJS..227....9P}, \cite{2013ApJ...775..127S}, \cite{2018MNRAS.475.2097C}). The rest of the paper is structured as follows: in section 2 we describe our detection and determination of beam corrections in the {\em AKARI} BSCv2 catalog; in section 3 we summarise the other data and fitting methods used with the corrected {\em AKARI} fluxes to fit model SEDs to the HERUS ULIRGs and present the results of these fits; in section 5 we discuss these results and further test our beam correction scheme by analysis using the closer and more extended {\em Herschel} Reference Sample galaxies \cite{2014MNRAS.440..942C}. Our conclusions are summarised in section 5. We assume a Hubble constant of $H_0 = 70$ kms$^{-1}$ Mpc$^{-1}$ and density parameters of $\Omega_M = 0.3$ and $\Omega_L = 0.7$.

\section{Beam Corrections for {\em AKARI}}

We search for evidence that beam corrections are needed for the {\em AKARI} BSCv2 catalogs (\cite{y18}) using the same method adopted by \cite{2017PKAS...32..293R} in their analysis of the {\em AKARI} BSCv1 catalog. The idea is to look for any correlation between a measure of the size of a source and the flux ratio between appropriate  {\em IRAS} and {\em AKARI} bands. We follow \cite{2017PKAS...32..293R} in using the difference between the 2MASS point source catalog and 2MASSX extended source $J$ band magnitudes as a measure of the extendedness of a source.

\begin{figure*}
\begin{tabular}{cc}
  \includegraphics[width=8.5cm]{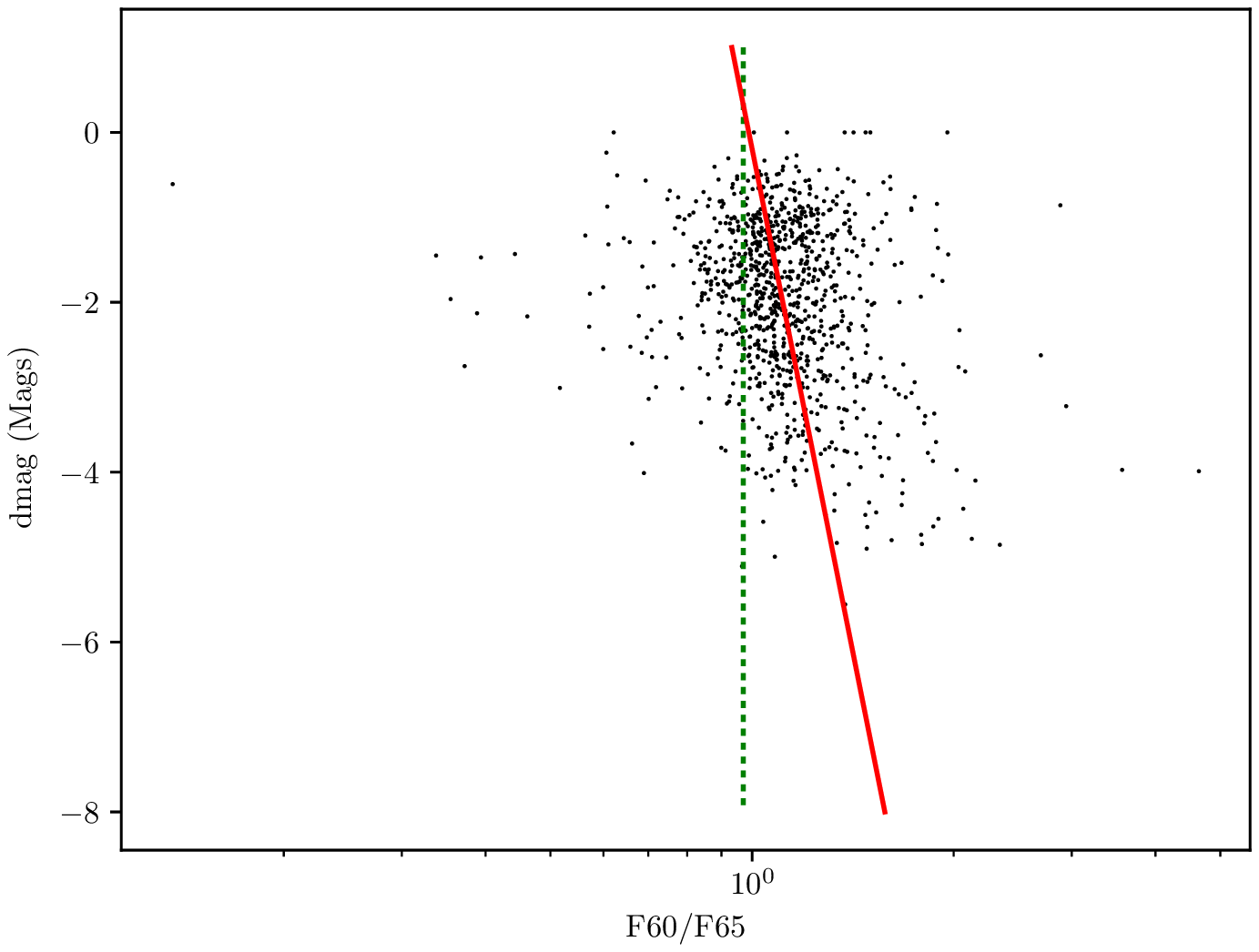} &
  \includegraphics[width=8.5cm]{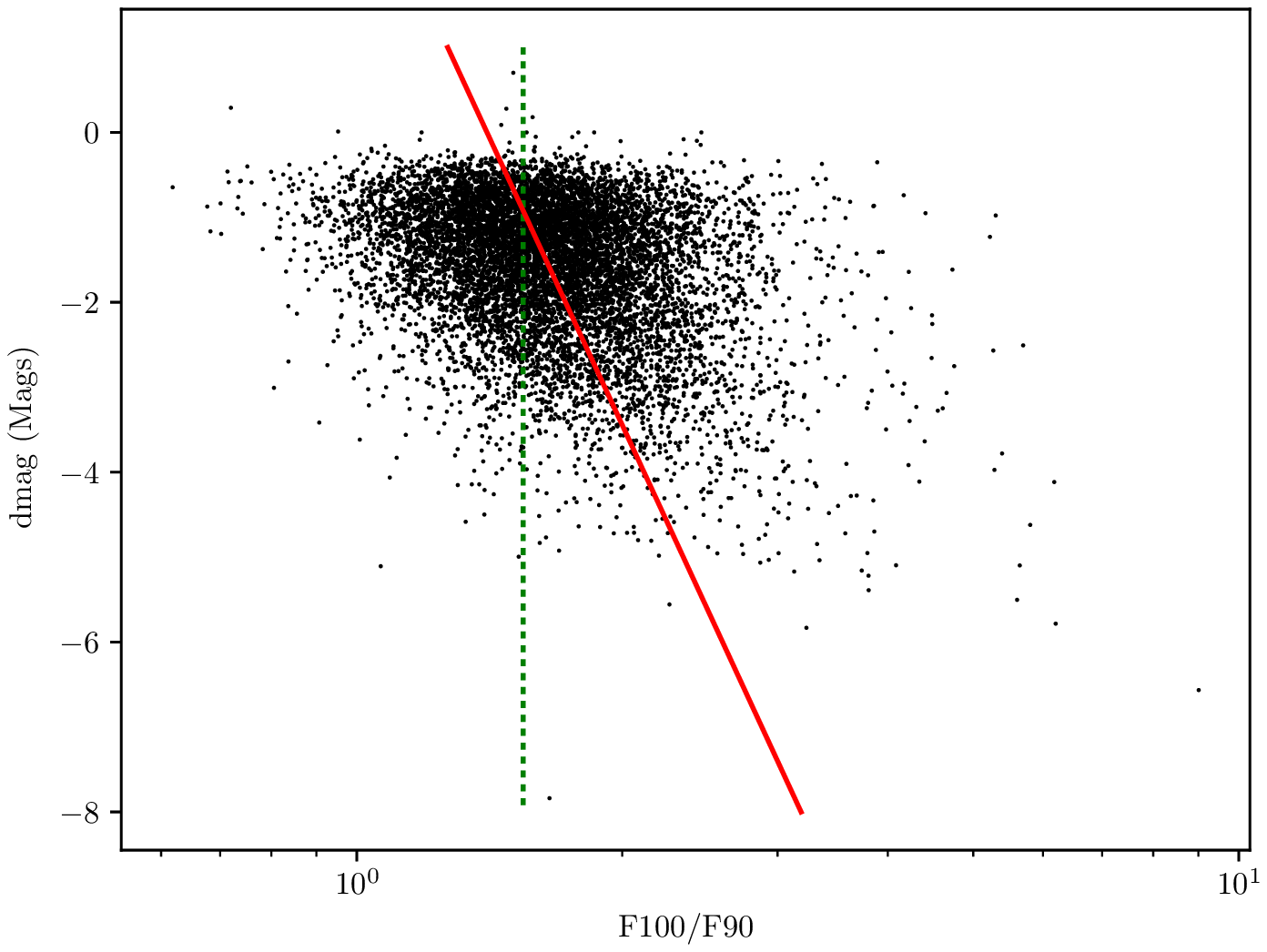} \\
    \includegraphics[width=8.5cm]{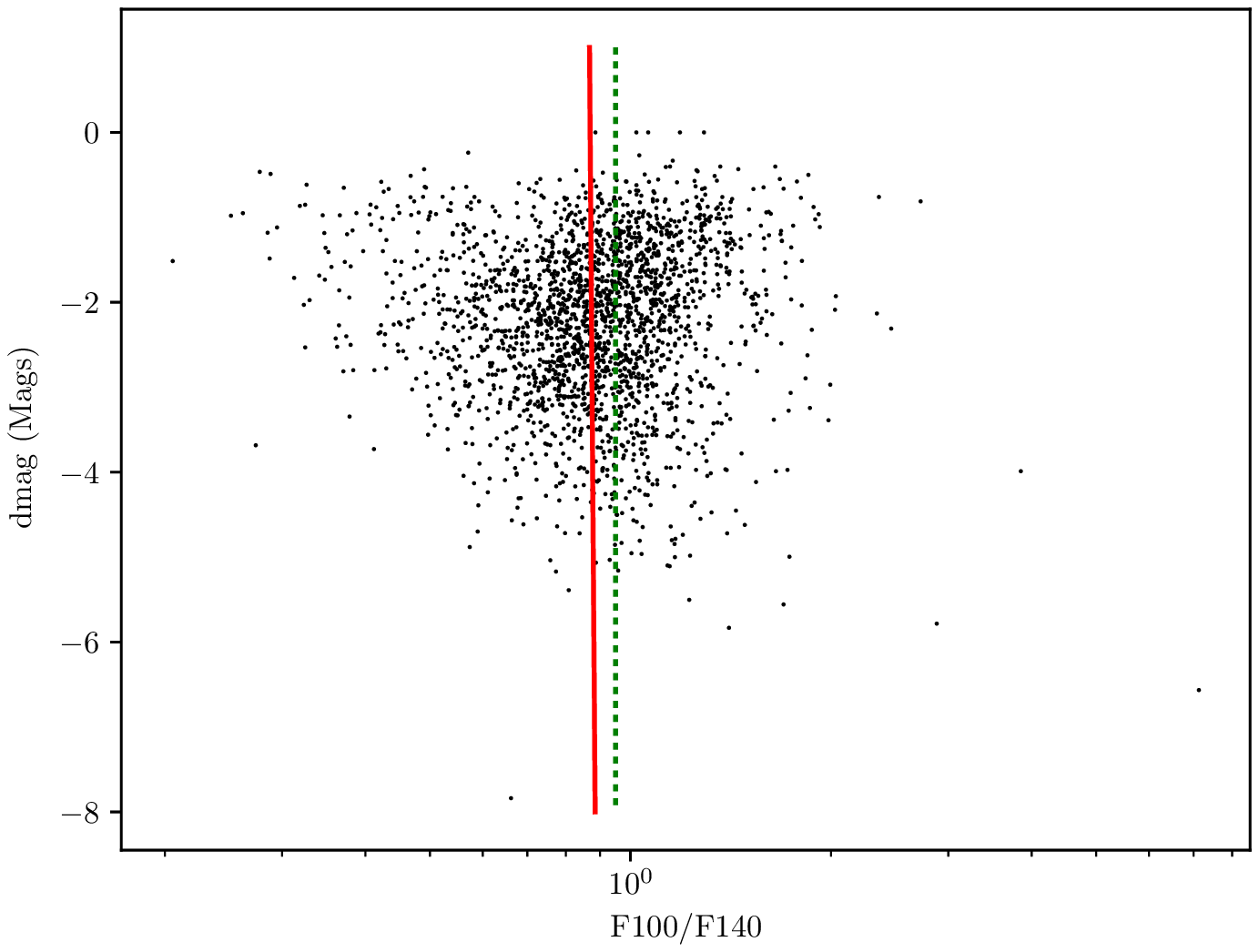} &
      \includegraphics[width=8.5cm]{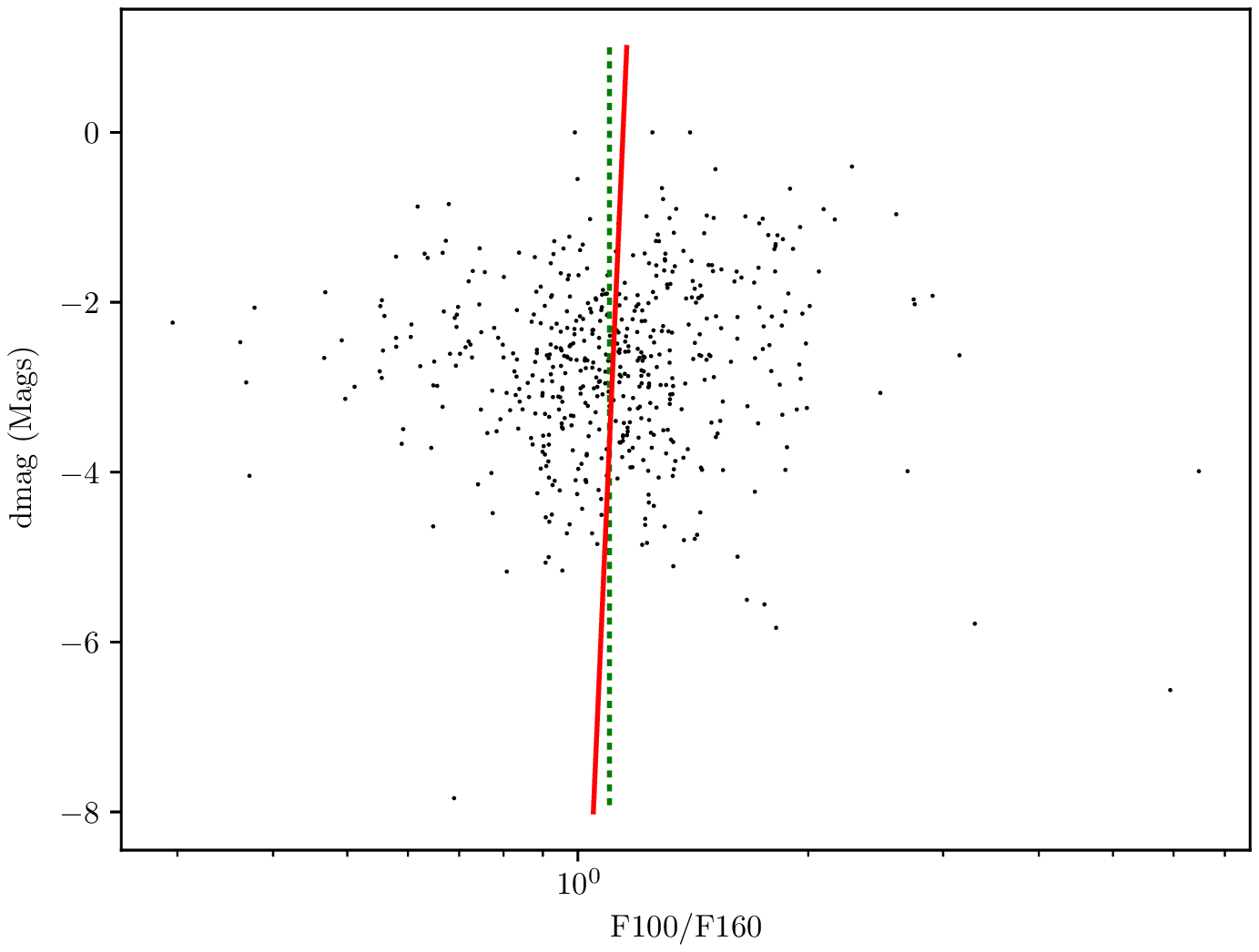} \\   
 \end{tabular}
\caption{The correlation between dmag, the difference between extended and point source J band 2MASS fluxes of our {\em AKARI} sources, and the {\em IRAS} to {\em AKARI} fluxes in the four bands. The solid red line shows the line of best fit for these correlations. The green dotted line indicates the relation that would be seen if there were no aperture correction for a single temperature $\beta=1.8$ modified black body at T= 45, 17, 25 and 25K for the 65, 90, 140 and 160$\mu$m plots respectively. Different selection effects pertaining to the different {\em IRAS} and {\em AKARI} bands mean that the galaxy populations appearing in these different plots are different, so that these notional `average' temperatures vary from band to band. $\beta$ of 1.8 is chosen since it is the median value for the {\em Herschel} Reference Sample of local galaxies \cite{2014MNRAS.440..942C}.}
\label{fig:corr}
\end{figure*}

We start by matching {\em IRAS} FSC sources to {\em AKARI} BSCv2 sources. We find there are 18,549 FSC sources that have BSCv2 counterparts. This compares with the 18,153 such sources found in a similar matching process by \cite{2017PKAS...32..293R} for the BSCv1 catalog. These sources are then matched to the 2MASS point source catalog and the 2MASS extended source catalog (2MASSX). A matching radius of 2 arcseconds is used to match 2MASS and 2MASSX sources, and 60 arcseconds for the {\em IRAS} to {\em AKARI} matches. We then calculate the difference in magnitude ($dmag$) between the 2MASSX extended and 2MASS point source J band magnitudes for these sources ie.
\begin{equation}
dmag = J_{2MASSX} - J_{2MASS}.
\end{equation}
This value represents a measure of the level of extension of the sources. We then calculate the ratio of {\em AKARI} and appropriate {\em IRAS} fluxes (F60/F65, F100/F90, F100/F140 and F100/F160) to see if there is any correlation between these ratios and $dmag$. The {\em IRAS} observations had larger beams than those from {\em AKARI} and will thus encompass the full flux of a source which some flux might be missed by {\em AKARI}. Any correlations between flux ratio and $dmag$ would indicate the need for a beam correction which can then be calculated from the correlation. For this analysis we restrict ourselves to those {\em AKARI} sources whose fluxes are of the highest quality, FQUAL=3, and which are detected at $>3\sigma$ in the appropriate {\em AKARI} band. We also exclude {\em IRAS} sources likely to be contaminated by cirrus emission (CIRRUS$>$1) or that have poor flux quality (FQUAL$>$2 for 60$\mu$m fluxes and FQUAL$>$1 for 100 $\mu$m fluxes). This restricts the number of sources  in our fit to 1493, 18451, 3508 and 845 in the 65, 90, 140 and 160$\mu$m bands respectively.

Plots of {\em IRAS} to {\em AKARI} flux ratio vs. $dmag$ are shown in Figure \ref{fig:corr}. There appear to be clear correlations between $dmag$ and flux ratio for the 65 and 90$\mu$m bands, while the presence of correlations in the 140 and 160$\mu$m bands is less clear. We fit lines to these correlations using a linear regression method (the \verb!stats.linregress! tool in \verb!scipy!) which also allows us to determine the likelihood that any correlation is real. The fitted lines are shown in Figure \ref{fig:corr}. The p-values from this calculation show that the $dmag$ correlations with the 90 and 65$\mu$m flux ratios are highly significant (chances of occurring at random are 5.7 $\times 10^{-250}$ and 6.8 $\times 10^{-14}$). Conversely the p-values show that there is no evidence for a correlation between $dmag$ and the 140 and 160$\mu$m flux ratios (chances of occurring at random are 0.77 and 0.44 respectively).

We thus conclude that beam corrections are necessary to compare {\em AKARI} 90 and 65 $\mu$m fluxes. The corrections derived from the slope of these correlations are:

\begin{equation}
\Delta \log{F90} = -\left(0.045 \pm 0.001\right) dmag\\~\\
\Delta \log{F65} = -\left(0.025 \pm 0.003\right) dmag
\end{equation}

The correction at 90$\mu$m is somewhat smaller than the correction of $-0.06 dmag$ derived in \cite{2017PKAS...32..293R}. These corrections must be applied to {\em AKARI} fluxes to allow them to be compared to data from other observatories such as {\em IRAS}, {\em Herschel} and ground-based submm observations.

\section{Fitting ULIRG SEDs}

\begin{figure*}
\begin{tabular}{cc}
  \includegraphics[width=8.5cm]{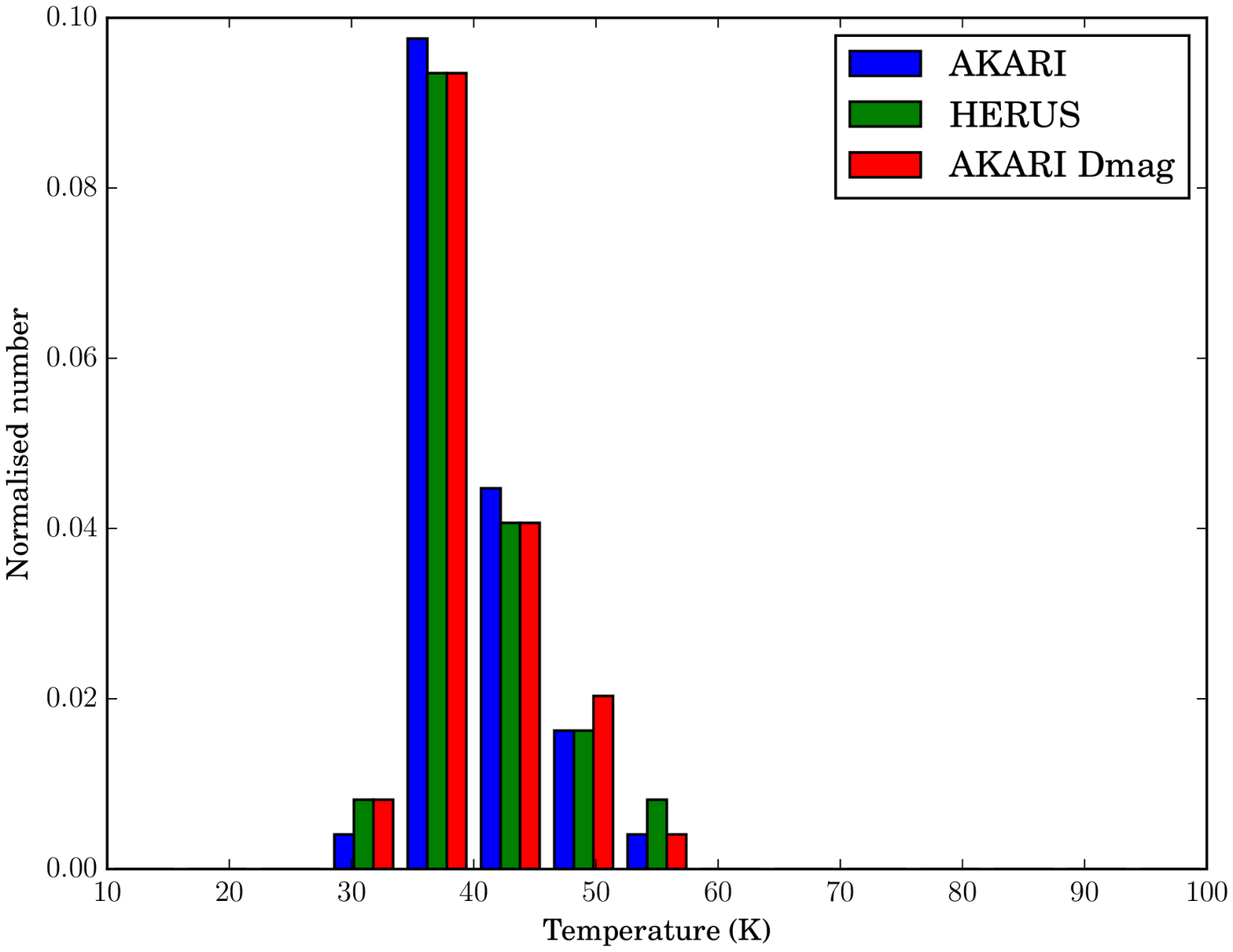} &
  \includegraphics[width=8.5cm]{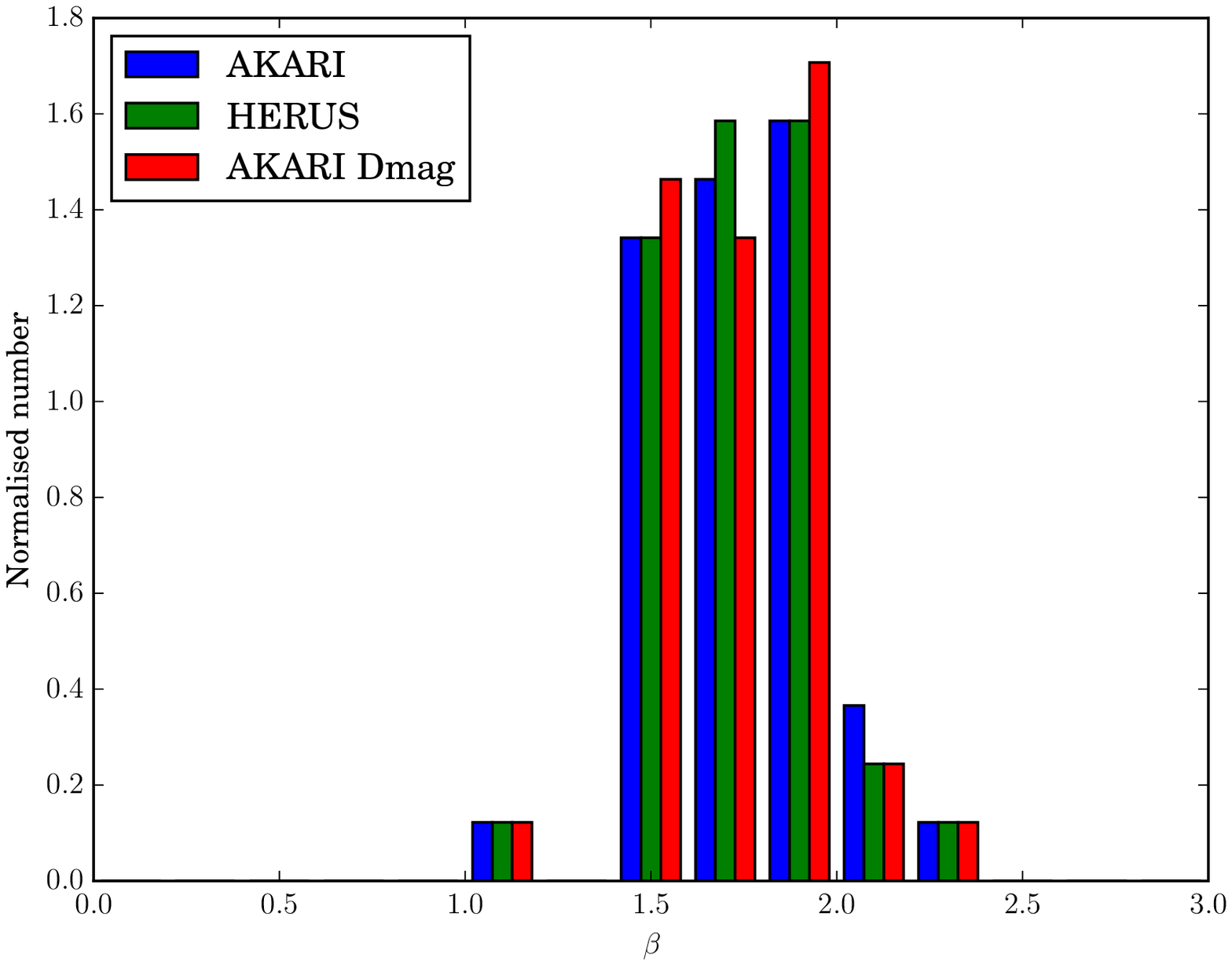} \\  
 \end{tabular}
\caption{The results of MBB fitting to the original {\em IRAS} and {\em Herschel} data from HERUS (green), to this data plus the {\em AKARI} without beam correction (blue) and the {\em AKARI} data after beam correction (red). We show histograms of both the derived temperature (left) and $\beta$ (right) distributions.}
\label{fig:fits}
\end{figure*}

\begin{figure*}
\begin{tabular}{cc}
\includegraphics[width=8.5cm]{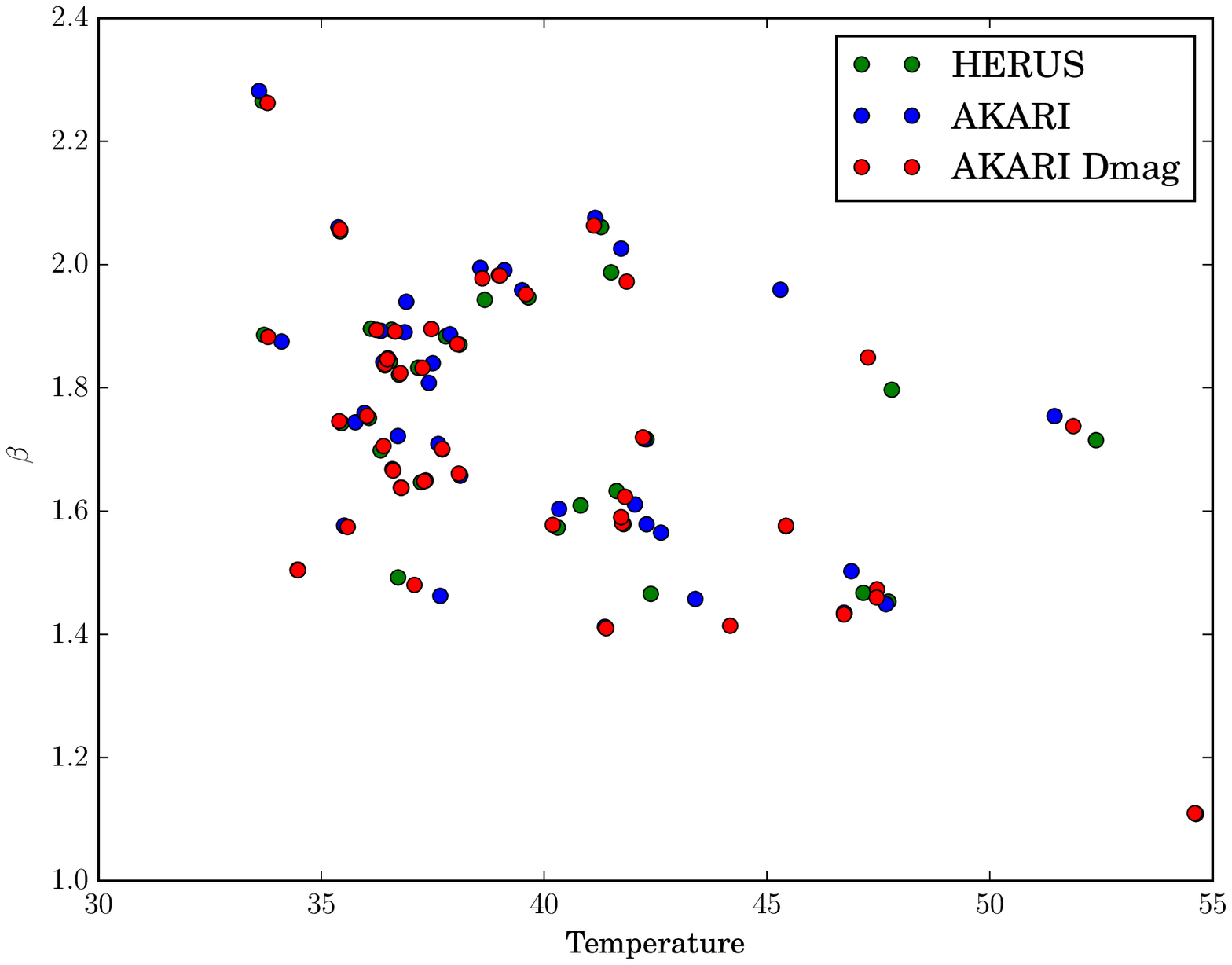} &
\includegraphics[width=8.5cm]{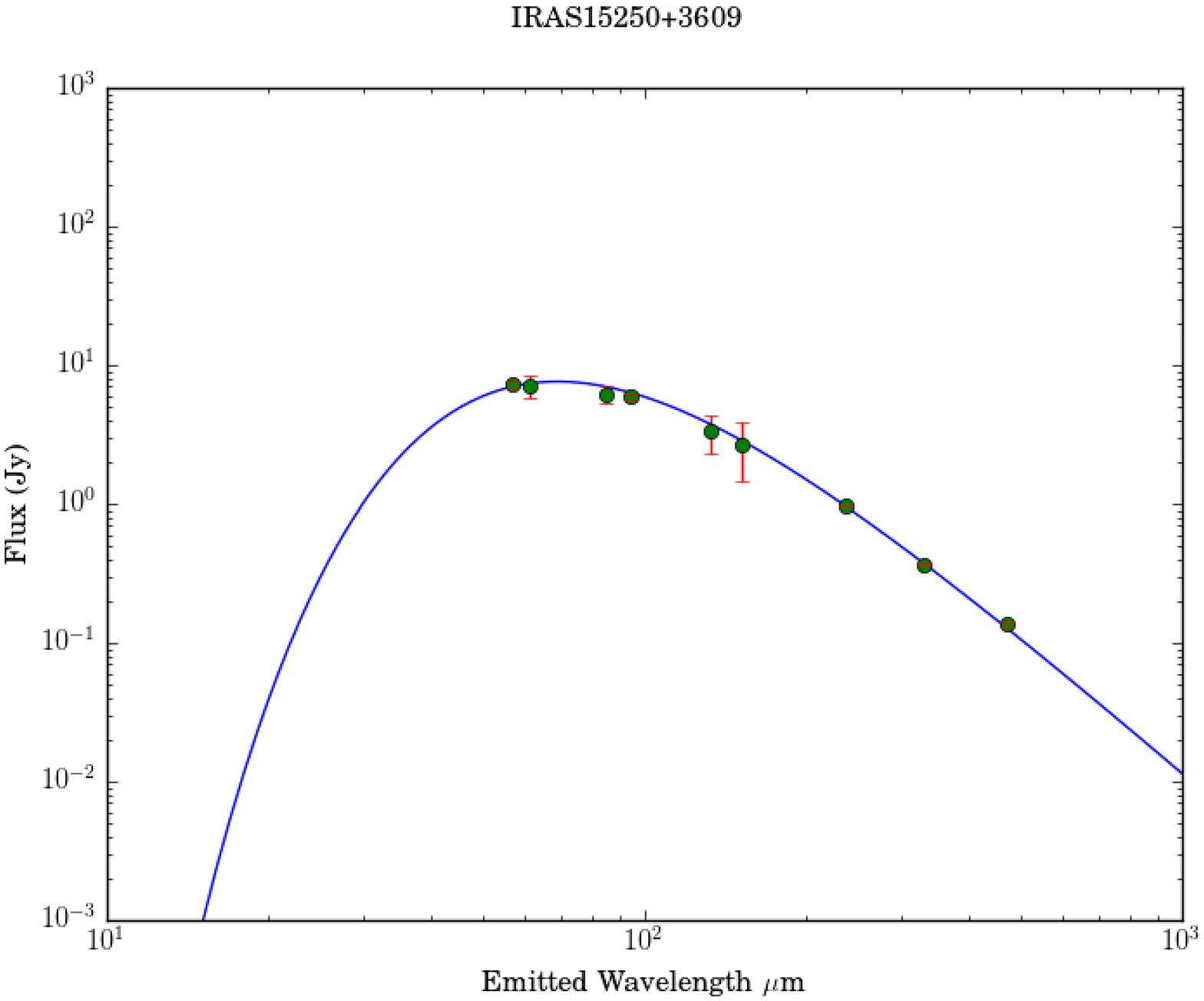}
\end{tabular}
\caption{Left: The temperature-$\beta$ relation for the original fits and for the fits including {\em AKARI} data both with and without beam correction. Colours as in Figure \ref{fig:fits}. Right: An example SED fit including the beam corrected {\em AKARI} fluxes, for the ULIRG IRAS15250+3609.}
\label{fig:tb}
\end{figure*}

{\em AKARI} data has the potential to make significant contributions to our knowledge of the far-IR SEDs of galaxies. We examine the effects our aperture flux corrections
might make to such SED fits by using the HERUS ({\em Herschel} ULIRG Survey \cite{2013ApJ...776...38F}) ULIRGs as a test case. These sources are a complete sample of 41 local ULIRGs with $z<0.3$ and 60$\mu$m flux $>1.8$Jy. All the HERUS ULIRGs have fluxes at 60 and 100$\mu$m from {\em IRAS}, and 250, 350 and 500$\mu$m from {\em Herschel}. We have previously (\cite{2018MNRAS.475.2097C}) fitted modified black body (MBB) SEDs to these sources, where
\begin{equation}
F_{\nu}[\nu, T, \beta] \propto \nu^{\beta} B_{\nu}(\nu, T) 
\end{equation}
using a Bayesian Markov Chain Monte Carlo (MCMC) fitting method (see \cite{2018MNRAS.475.2097C} for details). We here use the same fitting method but add the {\em AKARI} data, both with and without the aperture correction, to the set of data to be fitted for each object. The results of these fits, showing histograms of the derived values of $T$ and $\beta$ are shown in Figure \ref{fig:fits} and the $(T,\beta)$ correlation shown in Figure \ref{fig:tb}.

As can be seen the distributions of $T$, $\beta$ and the $T-\beta$ correlation with and without the {\em AKARI} data are fairly similar. This would suggest that the {\em AKARI} data are broadly consistent with the fluxes from {\em IRAS} and {\em Herschel} for these ULIRGs. The SED fit also looks reasonable which is a necessary confirmation that the fitting process has worked properly. The reduced $\chi^2$ values of the fits confirm that the fitting is doing a good job, with median values of 2.22 for the original fits, with two degrees of freedom, and 1.4 for the fits using aperture corrected {\em AKARI} data, with 5 degrees of freedom. [For the non-corrected {\em AKARI} data the median reduced $\chi^2$ is 1.6]. There are, however, long tails to the distribution of $\chi^2$ values, suggesting either some further corrections to the {\em AKARI} data need to be applied in some cases, and/or that a simple MBB fit is inadequate for some sources.

\section{Discussion}

\subsection{More Complex ULIRG SEDs}

Our determination of the beam corrections needed at 65 and 90$\mu$m has allowed the {\em AKARI} BSCv2 fluxes to be incorporated into our ULIRG SED fitting project for HERUS. The additional data has not significantly changed the result of these SED fits for simple MBB models. However, the true power of the additional data is that it allows more complex and physically realistic models to be fitted to the observations of individual sources while retaining a good number of degrees of freedom to the fit. We attempt to do this for an optically thick dust model for the current dataset. For dust that becomes optically thick at a frequency of $\nu_0$ this yields an SED given by:
\begin{equation}
F_{\nu}[\nu, T, \beta] \propto B_{\nu}(\nu, T)  \left( 1-e^{-\tau} \right); \\~ \tau = \left( \frac{\nu}{\nu_0}\right)^{\beta}.
\end{equation}
This model SED adds one extra parameter, $\nu_0$ to the fits.

We fit this model to the {\em IRAS}, beam corrected {\em AKARI} and SPIRE data for the HERUS ULIRGs using the same MCMC fitting method as before. This results in posterior probabilities on the parameter values that are rather more complex than the simple single peaked distributions found for the simple $(T, \beta)$ models. For the optically thick models these distributions are typically double peaked, indicating that the SEDs are either optically thin with parameters comparable to those found in the MBB fits, or are optically thick typically with higher dust temperatures and in some cases with values of $\beta$ closer to 2. We can confirm that Arp220 appears to be best fitted with an optically thick dust SED with a temperature of $\sim$ 60K, and $\beta$ of $\sim$1.7 and dust that becomes optically thick at $\sim$250$\mu$m. These results are very similar to those derived for this object by Rangwala et al. (2011) using a mix of {\em IRAS}, {\em Herschel}, {\em Planck}, {\em ISO} and SCUBA data. Other HERUS sources where there appears to be a strong case for optically thick dust in our fits include UGC5101, Mrk231, IRAS20087-0308, IRAS07598+6508 and IRAS16090-0139. Further analysis of these fits and the addition of additional fluxes from {\em Herschel}, {\em ISO}, SCUBA and other instruments for these sources is underway. This all demonstrates the potential power of combining {\em AKARI} data with the existing {\em IRAS} and {\em Herschel} data sets for which the beam corrections calculated here are an essential part.

\subsection{Beyond ULIRGs: The {\em Herschel} Reference Sample}

While the study of ULIRG SEDs was our original motivation for matching {\em AKARI} data to {\em IRAS} and {\em Herschel} photometry, it can be argued that they are not the ideal set of targets for testing a proposed beam correction method  since they are fairly compact sources with limited angular extent. We therefore further test our beam correction method using the {\em Herschel} Reference Sample (HRS, \cite{2014MNRAS.440..942C}). The HRS is a volume limited sample of 323 local galaxies selected to lie at distances between 15 and 25 MPc and to have $K<12$ for spiral and irregular types and $K<8.7$ for ellipticals and lenticulars. We cross matched the HRS catalog with the {\em AKARI} BSCv2 catalog using a matching radius of 15 arcseconds and further restricted this comparison sample to sources that had good, ie. $FQUAL = 3$, fluxes in all {\em AKARI} bands. This produced a sample of 41 galaxies with semi-major axis, as measured by {\em Herschel}, from 71 to 304 arcseconds. These sources are then matched to the 2MASS catalogs to allow $dmag$ values to be derived.

In addition to the SPIRE fluxes at 250, 350 and 500$\mu$m, the HRS galaxies also have PACS fluxes at 100 and 160$\mu$m. SEDs were fit to this data by the HRS team, including data from 100 to 500$\mu$m \cite{2014MNRAS.440..942C}, using a $(T,\beta)$ model, with successful fits found for 34 of the comparison sample of 41 objects (see \cite{2014MNRAS.440..942C} for details). These fits did not include a flux point shortward of 100$\mu$m.

We similarly fit SEDs to these sources with $(T, \beta )$ models using the methods described above. For this comparison we used several selections of data:

\begin{itemize}

\item Firstly we did a direct comparison to the results of \cite{2014MNRAS.440..942C} by fitting SEDs to the PACS and SPIRE data with the addition of the {\em AKARI} 140 and 90$\mu$ (corrected) fluxes. This comparison did not include the 65 or 160 $\mu$m fluxes from {\em AKARI}. The resulting SEDs generally match the temperature and beta values from \cite{2014MNRAS.440..942C} well within the errors on the fits, though there is a tendency for the fits that include the {\em AKARI} data to have slightly lower derived temperatures by about 1K. There are three sources that have more significant differences in fitted temperature, ranging from $\sim$ 1.5 to 2$\sigma$ significance. Inspection of these fitted SEDs shows that these three sources all have their {\em AKARI} 140$\mu$m flux significantly low in comparison to the PACS 100 and 160$\mu$m fluxes. They are also among the larger HRS sources, suggesting that there may need to be some additional beam correction that our analysis using 2MASS $J$ band fluxes has not managed to detect.

\item Secondly we replace the PACS 160$\mu$m flux with the {\em AKARI} flux in this band and repeat the fits above. This produces fits that are broadly consistent with the HRS fits using just PACS and SPIRE data. However, inspection of the fitted SEDs shows that in many cases the {\em AKARI} 160$\mu$m point is significantly below the fitted model. Comparison of the PACS and {\em AKARI} flux measurements at 160$\mu$m shows that the PACS fluxes are significantly larger than the {\em AKARI} fluxes, with flux ratios ranging from 1.15 to 3.75 with the significance of these differences ranging from 2 to 10 $\sigma$. In general the greatest discrepancies come from the largest galaxies, arguing that beam corrections are needed at 160$\mu$m for sources that are several arcminutes in size. One possible explanation for this problem is that large sources might be `shredded', whereby a bright extended source is broken up into several separate smaller, fainter sources in the catalog. Examination of the larger HRS sources finds no evidence for shredding in the BSCv2 catalog or in the FISv1 images.

\item Finally we examine the SED fits using fluxes from SPIRE at 250, 350 and 500$\mu$m, PACS at 100 and 160$\mu$m, {\em AKARI} at 140 and 90$\mu$m (corrected), and adding the corrected {\em AKARI} 65$\mu$m and {\em IRAS} 60$\mu$m fluxes. With this set of data, the fits, compared to the original HRS fits using {\em Herschel} 100 to 500$\mu$m data \cite{2014MNRAS.440..942C}, are skewed to higher temperatures by about 4K, lower $\beta$ values by $\sim$ 0.5. Comparison of the fits to the data shows that the problem cases identified above continue to be problematic, but also that the {\em AKARI} and PACS fluxes at 90 and 100$\mu$m almost universally lie below the fit. This suggests that the 60 and 65$\mu$m points from {\em IRAS} and {\em AKARI} may well be contaminated with higher temperature dust and that the assumption of a single temperature fit to the dust is not valid. Further analysis of this issue is beyond the scope of the current paper.

\end{itemize}

From this analysis we can conclude a number of things. Firstly, that while the $dmag$ based beam corrections seem to work reasonably well for objects like ULIRGs with little spatial extension, for the larger HRS galaxies these corrections. Since the HRS galaxies are large sources with differing inclinations and detailed dust distribution it is not surprising that a correction based simply on the ratio of 2MASS $J$-band point and extended magnitudes is incomplete. The majority of the 18549 sources used in the determination of the $dmag$ beam correction are quiet small in extent, with fewer than 1\% having sizes comparable to the most problematic galaxies in the HRS comparison. The mean Kron semi-major axis for the 18549 sources, as given in the 2MASS Extended Source Catalog, is 19.4 arcseconds (median is 13.6 arcseconds). For the HRS sources it is 73 arcseconds (median 59 arcseconds) so larger and more complex beam correction effects might well be expected.

Secondly, comparison of {\em AKARI} 140 and 160$\mu$m data to the fits and PACS 160$\mu$m fluxes for the HRS sources suggests that beam corrections are needed for the 140 and 160$\mu$m bands are needed for these sources. The PSF for these bands has a FWHM of 58 and 61 arcseconds respectively at 140 and 160$\mu$m respectively \cite{2007PASJ...59S.389K}, so it is not surprising that some flux is lost in observations of galaxies with a mean semi-major axis of 73 arcseconds.

We conclude that our $dmag$-based beam correction method is appropriate for moderately extended objects like the local ULIRGs, but that for more extended objects, such as the nearby galaxies in the HRS, a more sophisticated approach is required that can deal with the detailed structure of each individual source. Whether this can be done in the context of the BSC or if an additional product, along the lines of the {\em IRAS} Small Scale Structure Survey is not clear.

\section{Conclusions}

We have derived beam corrections for {\em AKARI} BSCv2 sources by correlating the ratio of {\em AKARI} fluxes to {\em IRAS} fluxes, in appropriate bands, to the difference in magnitude in the $J$ band between the 2MASS point source and extended source catalogs. We find that beam corrections are needed in both the 65 and 90$\mu$m bands, but no corrections are needed for the 140 and 160$\mu$m bands, which have larger beams on the sky. We then compared the results of simple MBB SED fitting models to the HERUS ULIRGs derived from {\em IRAS} and {\em Herschel} data alone and those found when combining this with corrected and uncorrected {\em AKARI} data. For the simple MBB fits we find no significant change in the SED properties, but the reduced $\chi^2$ values are best for the dataset that includes the beam corrected {\em AKARI} data. We also attempt to fit an optically thick SED model. We find that in most cases this yields degenerate fits that cannot distinguish between solutions that are optically thick or optically thin at far-IR wavelengths. However, for some of our ULIRGs the optically thick models appear to be favoured. This includes Arp220, for which we recover optically thick dust SED parameters very similar to those found by \cite{2011ApJ...743...94R}. This demonstrates the value of including the beam-corrected {\em AKARI} data in such studies. We also test our beam correction method on the {\em Herschel} Reference Sample galaxies \cite{2014MNRAS.440..942C} which have larger angular extents than those of the HERUS ULIRGs or the majority of the 2MASS sources used in our beam correction measurement. For these sources we find evidence that additional beam corrections are needed at 140 and 160$\mu$m and that our simple approach used for beam correction at 90 and 65$\mu$m may not be sufficient. For nearby galaxies, extended on scales of 1 arcminute or more, specific extended source processing appears to be be needed beyond what currently exists in the {\em AKARI} pipeline. Improvements to the pipeline to allow the full recovery of extended source fluxes, or the possible addition of a Small Extended Source Catalog would significantly enhace the usefulness of the {\em AKARI} data in the context of multiwavelength and multimission photometric studies. A reduction in the currently quite large calibration errors at 140 and 160$\mu$m  would also be very helpful for such projects.


%


\begin{ack}
This research is based on observations with {\em AKARI}, a JAXA project with the participation of ESA.
SPIRE has been developed by a consortium of institutes led
by Cardiff Univ. (UK) and including: Univ. Lethbridge (Canada);
NAOC (China); CEA, LAM (France); IFSI, Univ. Padua (Italy);
IAC (Spain); Stockholm Observatory (Sweden); Imperial College
London, RAL, UCL-MSSL, UKATC, Univ. Sussex (UK); and Caltech,
JPL, NHSC, Univ. Colorado (USA). This development has been
supported by national funding agencies: CSA (Canada); NAOC
(China); CEA, CNES, CNRS (France); ASI (Italy); MCINN (Spain);
SNSB (Sweden); STFC, UKSA (UK); and NASA (USA). This publication makes use of data products from the Two Micron All Sky Survey, which is a joint project of the University of Massachusetts and the Infrared Processing and Analysis Center/California Institute of Technology, funded by the National Aeronautics and Space Administration and the National Science Foundation. E.GA is a Research Associate at the Harvard-Smithsonian
Center for Astrophysics, and thanks the Spanish 
Ministerio de Econom\'{\i}a y Competitividad for support under projects
FIS2012-39162-C06-01 and  ESP2015-65597-C4-1-R, and NASA grant ADAP
NNX15AE56G. JBS wishes to acknowledge the support of a Career Integration Grant within the 7th European Community Framework Program, FP7-PEOPLE-2013-CIG-630861-FEASTFUL. J.A. acknowledges financial support from the Science and Technology Foundation (FCT, Portugal) through research grants PTDC/FIS-AST/2194/2012 and UID/FIS/04434/2013.
DLC and JG acknowledge support from STFC, in part through grant numbers ST/N000838/1 and ST/K001051/1. The authors wish to thank A. Jaffe for useful conversations, and DLC would like to thank the {\em AKARI} project, Tokyo University and K. Kohno for their generous hospitality.

\end{ack}

%
%


\end{document}